\documentclass{emulateapj}

\usepackage{epsfig}

\newcommand{\be}{\begin{equation}}
\newcommand{\ee}{\end{equation}}
\newcommand{\ba}{\begin{eqnarray}}
\newcommand{\ea}{\end{eqnarray}}

\shorttitle{Primordial Power Spectrum Reconstruction}
\shortauthors{Mukherjee \& Wang}

\begin{document}

\title{Primordial Power Spectrum Reconstruction}
\author{Pia~Mukherjee$^{1,2}$ and Yun~Wang$^1$}
\affil {$^1$ Department of Physics \& Astronomy, University of Oklahoma,
                 440 W Brooks St., Norman, OK 73019, USA;
                 email: pia,wang@nhn.ou.edu}
\affil {$^2$ Astronomy Centre, University of Sussex, Brighton BN1 9QH, UK}

%\altaffiltext{1}{Present address: Astronomy Center, University of Sussex, Brighton BN1 9QH, UK}

\begin{abstract}

In order to reconstruct the initial conditions of the universe it is
 important to devise a method that can efficiently constrain the shape of 
the power spectrum of primordial matter density fluctuations in a 
model-independent way from data. In an earlier paper we proposed a method
 based on the wavelet expansion of the primordial power spectrum. 
The advantage of this method is that the orthogonality
 and multiresolution properties of wavelet basis functions enable 
information regarding the shape of $P_{\rm in}(k)$ to be encoded in a small 
number of non-zero coefficients. Any deviation from scale-invariance can then
 be easily picked out. Here we apply
 this method to simulated data to demonstrate that it can accurately
reconstruct an input $P_{\rm in}(k)$, and present a prescription for how this 
method should be used on future data.

\end{abstract}


\section{Introduction}

Cosmic microwave background (CMB) and large scale structure (LSS) data are powerful probes 
of the universe. They are sensitive not only to the values of cosmological parameters but 
also to the initial conditions of the universe that may have arisen due to inflation. 
While the general concept of inflation solves many problems in cosmology and provides a 
framework within which the data are very well fit, not much is known about the details 
of inflation. 

The power spectrum of density perturbations that results from inflation is usually assumed to be 
scale-invariant, or sometimes a power-law or a power-law with a running tilt, and such 
parametrizations fit the data well so far. Simple models of inflation generically predict 
Gaussian adiabatic perturbations with a nearly scale-invariant spectrum. However there are 
other possibilities, for example if multiple scalar fields were present during inflation, 
or if the slow-rolledness of the inflaton field was temporarily interrupted due to a 
feature in the inflaton potential, etc. Such models can predict power spectra that are 
different from scale-invariance/power-law 
\citep{Holman91ab,Wang94,Randall96,Adams97,lesgourgues00,Chung00,Adams01,Lyth02,FengZhang03,Burgess05}.

Since the details of inflation are unclear, we could use precision observations 
of the CMB and LSS to constrain not only the cosmological parameters but also the 
shape of the primordial power spectrum ($P_{\rm in}(k)$) in a non-parametric way. The 
motivation is to test the assumptions that are usually made regarding the shape
of $P_{\rm in}(k)$, and to check if there are any features in $P_{\rm in}(k)$ 
that can thus be unravelled. The aim 
is to try to see if, as the data continue to improve, something new can be learnt 
about interesting cosmologically relevant functions such as $P_{\rm in}(k)$ besides 
deriving stronger constraints on cosmological parameters based on the same set of assumptions.  
We presented an efficient method for reconstructing the shape of $P_{\rm in}(k)$ in 
Mukherjee \& Wang (2003c). In this paper, we demonstrate 
that this method is able to accurately reconstruct an input 
$P_{\rm in}(k)$ by applying it to simulated data. 

That $P_{\rm in}(k)$ be probed model-independently was initially suggested by 
\cite{Wang99} and \cite{WangMathews02}, who used conventional top-hat binning 
and linear interpolation
 respectively in the parametrization of $P_{\rm in}(k)$. Follow up analysis using pre-WMAP data 
 and WMAP data was done by Mukherjee \& Wang (2003a) and \cite{pia03b}. In Mukherjee \& 
 Wang (2003c) we proposed the wavelet expansion method and used it on WMAP data. This method 
 is superior to binning methods in that it has less correlated errors. The method also 
 has various other advantages: due to the simultaneous and adaptive localization property 
 of orthogonal wavelets in both scale and position spaces, the information content 
 of most signals is sparsely represented in wavelet space with useful information being 
 encoded only in a small number of coefficients that are large, while the others are small, 
 or in the case of noisy data, consistent with zero. One can then hope to recover the large 
 coefficients more easily and thus constrain the function more efficiently. Function 
 reconstruction and denoising are thus efficiently achieved with wavelets, as we demonstrate 
 further in this paper. 
 
 Since complex projection effects and other degeneracies limit 
 our ability the constrain a function in $k$ space using data in multipole space 
 (Tegmark \& Zaldarriaga 2002), the use of a 
 different basis set, that is at the same time adaptive to the signal at hand, 
 is certainly a useful complementary method to any other model-independent method. 
 Since the release of the WMAP first year data (Spergel et al. 2003, Peiris 
et al. 2003), a number of analyses with emphasis on the primordial power spectrum 
have been undertaken (Bridle et al. (2003), Barger, Lee \& Marfatia (2003), 
Seljak, McDonald \& Makarov (2003)). Model-independent methods have been pursued 
in Bridle et al. (2003), \cite{hannestad04,Hu04,matsumiya02,SS04,valentini04,kogo04}. 

In this paper, we use simulated data to explicitly demonstrate the potential and reliability 
of the wavelet expansion method in primordial power spectrum reconstruction.
Sec.2 describes our method. Sec.3 contains results. We conclude in Sec.4.

\section{Method}

CMB data are sensitive to $P_{\rm in}(k)$ as follows:
\begin{equation}
C_l(\{b_{j,l}\})=(4\pi)^2 \int \frac{dk}{k} P_{\rm in}(k) 
\left|\Delta_{Tl}(k, \tau=\tau_0)\right|^2
\end{equation}
where the cosmological model dependent transfer function 
$\Delta_{Tl}(k,\tau=\tau_0)$ is an integral over
 conformal time $\tau$ of the sources which generate CMB 
 temperature fluctuations,
 $\tau_0$ being the conformal time today.

As described in \cite{MW03c} we parametrize $P_{\rm in}(k)$ by
the coefficients of its wavelet expansion.
\begin{equation}
P_{\rm in}(k_i) = \sum_{j=0}^{J-1}\sum_{l=0}^{2^j-1} b_{j,l}\psi_{j,l}(k_i),
\label{WT}
\end{equation}
where $\psi_{j,l}$ are the wavelet basis functions,
constructed from the dilations and translations of a  
mother function $\psi(k)$.
The resulting wavelet bases are discrete,
compactly supported and orthogonal with respect to both the scale $j$ 
and the position $l$ indices, and hence their coefficients provide a 
complete and non-redundant (hence invertible) representation of the
 function $P_{\rm in}(k)$.
The scale index $j$ increases from 0 to $J-1$, and
 wavelet coefficients with increasing 
$j$ represent structure in $P_{\rm in}(k)$ on
increasingly smaller scales, with each scale a factor of 2 finer than
the previous one.  The index $l$, which runs from 0 to $2^j-1$ for each $j$, 
denotes the position of the wavelet basis $\psi_{j,l}$ within the $j$th
scale. (Note that the properties of the basis functions are completely
 different from the basis functions in binning methods). 

We take the sample points $k_i$ to be equally spaced
in $\log(k)$, and use 16 wavelet coefficients to represent $P_{\rm in}(k)$
in the range  $0.0001  \la k /(\mbox{Mpc}^{-1}) \la 0.1$, with the spectrum
 for $k<0.0001$ Mpc$^{-1}$ and $k>0.1$ Mpc$^{-1}$ set to be scale-invariant. 
We use the Daubachies-4 wavelet \footnote{The number associated with the wavelet 
corresponds to the order of the wavelet; the larger the number the smoother 
the wavelet is and the better it is at representing smooth functions or higher 
order polynomials.}
here \citep{Daub92,Press94}. In general different 
discrete orthogonal wavelets will give similar results.

We obtain constraints on the wavelet coefficients of $P_{\rm in}(k)$, as well
 as the usual cosmological parameters, from simulated data, using a Markov
 Chain Monte Carlo (MCMC) technique to trace the full posterior probability 
distribution of all the parameters. We use CAMB (appropriately modified) and 
CosmoMC for our computations. We simulate cosmic variance limited CMB
 temperature power spectrum data in a flat $\Lambda$CDM universe with 
 $H_0=70\,$km$\,$s$^{-1}$Mpc$^{-1}$, $\Omega_bh^2=0.022$, $\Omega_ch^2=0.12$ 
 and $\tau=0$, out to an $l_{max}$ of 2000, with some functional form for the
primordial power spectrum.  The forms we have considered are
 given in the next section, each normalized to $1.8\times10^{-9}$.
 The samples of coefficients 
can be mapped onto corresponding power spectra and the confidence limits on the 
power spectra can be obtained at each $k_i$, or equivalently constraints
 on the wavelet coefficients can be inverted to obtain constraints on the
 power spectrum with inter-correlations taken into account.

\section{Results}

We consider three different functional forms for the input $P_{\rm in}(k)$, the scale-invariant, 
saw-tooth, and Gaussian dip models.
The scale-invariant is given by 
\be
P_{\rm in}(k)=1.
\ee
The Gaussian dip model is given by
\be
P_{\rm in}(k)=1-\alpha e^{-\left[log(k/0.05)\right]^2/\beta}, 
\ee
where $\alpha=0.7$ and $\beta=0.1$.
The saw-tooth model is given by
\ba
P_{\rm in}(k) &=& a_1, \hskip 0.5cm k<k_1 \\
  &=& a_N, \hskip .5cm k>k_N   \nonumber\\ 
  &=& \frac{k_i-k}{k_i-k_{i-1}} a_{i-1} + 
\frac{k-k_{i-1}}{k_i-k_{i-1}} a_i, \hskip .5cm k_{i-1}<k<k_i,\nonumber
\ea
where $a_1=1$, $a_2=0.65$, $a_3=1.35$ and $a_N=1$, with the $k_i$'s 
stretching between 0.005 to 0.1 equally spaced in log(k). Note 
that these spectra have been selected independently of the discretization
 of the recovered power spectra. What the method efficiently fits for is the
 correspondingly discretized input spectra. 

The orthogonality of the wavelet basis functions in $k$ space is degraded 
due to the non-linear model-dependent mapping between $k$ and $l$ spaces
and due to degeneracies with cosmological parameters. As a result it is
 important to take into account the correlations between the constrained
 wavelet coefficients in deducing constraints on the primordial power 
spectrum. The correlations can be obtained with a sufficient number of MCMC
 samples. We obtained about $10^6$ samples for each case.
 
The $C_l$ spectra corresponding to these primordial power spectra are shown in Figure 1. Corresponding $C_l$ observations are then simulated for the WMAP and Planck satellite missions. For the WMAP experiment three frequency channels are used with beam FWHM of 31.8, 21.0 and 13.8 arcmins, and pixel noise of 19.8, 30.0, 45.6 $\mu$K, respectively. For the Planck experiment, three frequency channels with beam FWHM of 10.7, 8.0, and 5.5 arcmin, and pixel noise of 4.6, 5.4 and 11.7 $\mu$K, respectively, are used. For both experiments a sky fraction of 0.8 is assumed. Instrumental noise of these specifications together with cosmic variance uncertainties are used to simulate data (see for example Tegmark et al. 2000).

Figure 2 shows the input $P_{\rm in}(k)$ (dashed curve), and mean power 
spectrum reconstructed from simulated $C_l$ data together with 
2$\sigma$ uncertainties (solid curves). The dot-dashed curves give the 
5$\sigma$ contours. This is shown for the scale-invariant, 
saw-tooth and Gaussian dip models of $P_{\rm in}(k)$, for WMAP and Planck simulated observations of the CMB temperature anisotropy power spectrum. In each case the dashed curves shown are discretized versions of the corresponding continuous spectra. Figure 1 shows that the input $P_{\rm in}(k)$ is accurately recovered by our method. The features are detected at high significance. 

We have used CAMB with its original discretization in $k$
 and $l$ spaces with interpolation in between; this gridding can be made
 smoother to improve the accuracy of reconstructions if needed. 

When applied on future data, for example from the 4-year WMAP or Planck experiments, 
complemented by LSS power spectrum data, we recommend that our method be
used as follows:\\
(a) Follow the procedure described in this paper to deduce
 constraints on $P_{\rm in}(k)$. \\
(b) Look at constraints obtained from the larger (more significantly constrained) wavelet coefficients, i.e. those that deviate from zero at greater than say 1$\sigma$, or 2$\sigma$.
 If tentative evidence of a feature is found in the thus denoised $P_{\rm in}(k)$ spectrum, i.e. if any wavelet coefficients other than the two that together indicate/reconstruct the amplitude of a scale-invariant function are found to be significantly non-zero, then there is indication of a feature or a deviation from scale-invariance. \\
 (c) If step (b) finds the indication of a feature, one can use a
 parametric form for the kind of deviation from scale-invariance that is indicated, and repeat the likelihood analysis to better fit the feature and to obtain constraints on the relevant 
parameters. On the theoretical side, hopefully the feature will indicate something that is theoretically viable in certain 
scenarios so that it can be thus modelled. The relevant parametrization to use 
can also be derived in this way. If evidence calculations indicate that scale-invariance or power-law forms remain the
 best explanation for the data we will have achieved the task of checking
 for this model-independently and ruling out or constraining more complex physics
 within inflationary models. 

\section{Conclusions}

It is important to reconstruct the primordial power spectrum as a free function
from cosmological data, the aim being 
 to test the assumptions that are usually made about $P_{\rm in}(k)$,
 and to constrain it model-independently so as to be able
 to detect any features in it that may be signatures of new physics 
during inflation. Representing $P_{\rm in}(k)$ by coefficients of its wavelet expansion
is an effective and powerful way of approaching this problem as described above. We 
have demonstrated that input $P_{\rm in}(k)$ are accurately recovered with this method. We have 
presented a prescription for
 how this method should be used on future data.
 
We have used CMB temperature data simulated according to WMAP and Planck specifications, with various forms for the primordial power spectra. Our results (see Fig.2) indicate that future data hold great promise for probing the detailed shape of the primordial power spectrum.

\section{Acknowledgements}
We acknowledge the use of supercomputing facilities at the University of Oklahoma and Henry Neeman's help in making it accessible to us. We thank Andrew Liddle for comments on the manuscript. We acknowledge the use of CAMB and CosmoMC. This work is supported in part by NSF CAREER grant AST-0094335, and in part by PPARC.

\begin{figure}
\epsscale{0.6}\plotone{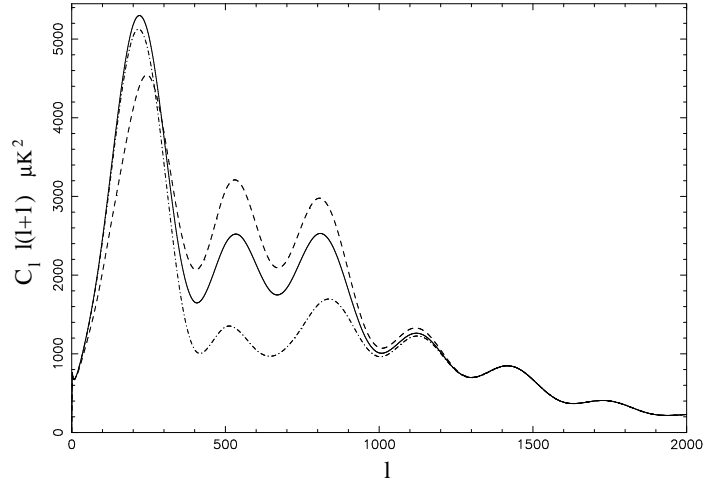}
\figcaption{The plot shows the temperature anisotropy $C_l$ spectrum resulting from a scale-invariant (solid curve), saw-tooth (dashed surve) and Gaussian dip (dot dashed curve) primordial power spectra.}
\end{figure}

\begin{figure}
\epsscale{0.9}\plotone{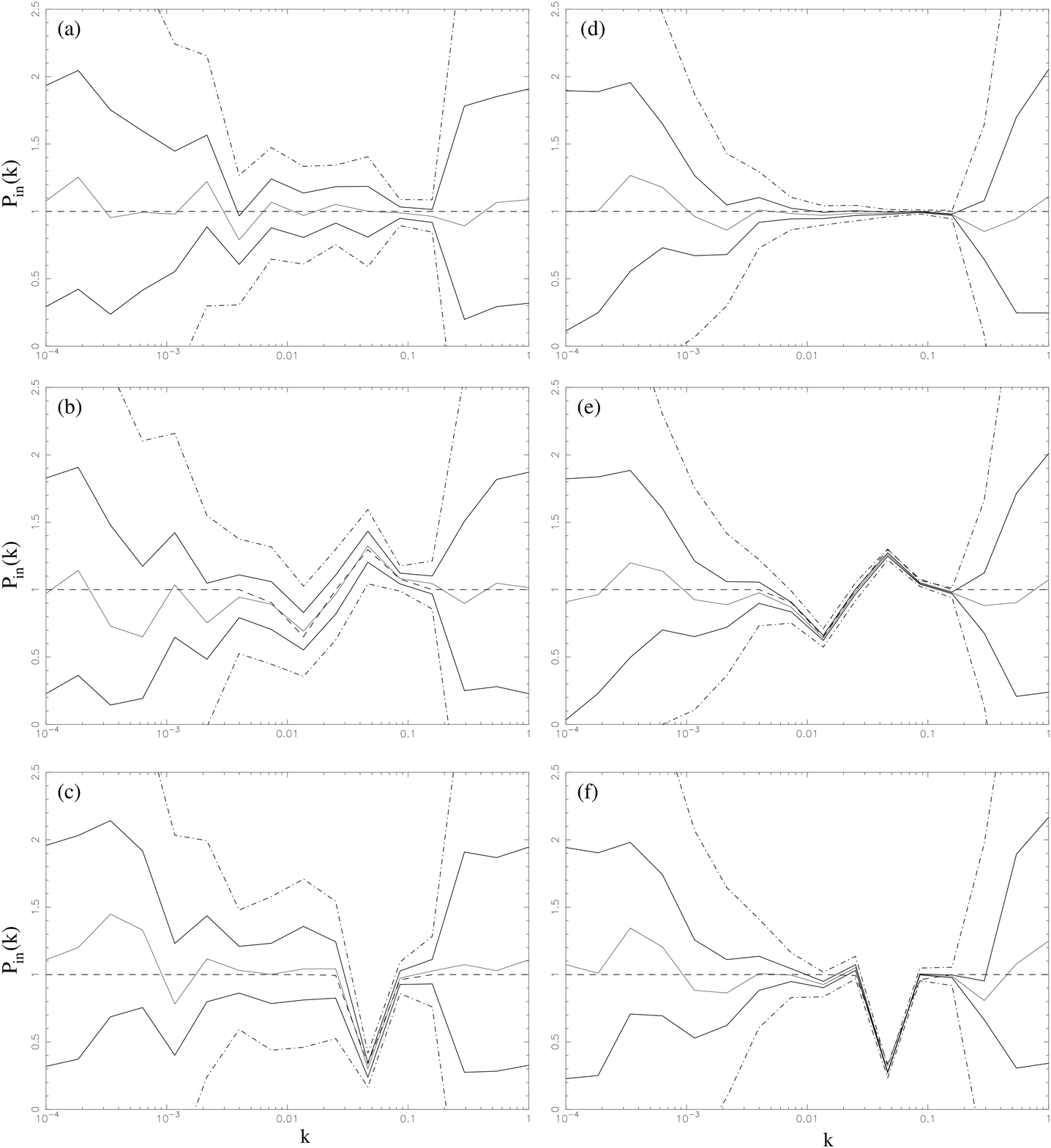}
\figcaption{Each plot shows the input primordial power spectrum (dashed curve), and mean power spectrum reconstructed from simulated $C_l$ data together with 2$\sigma$ uncertainties (solid curves) and the 5$\sigma$ confidence contours (dot-dashed curves). These are shown for WMAP simulated observations in the (a) scale-invariant, (b) saw-tooth and (c) Gaussian dip cases, and similarly for Planck simulated observations in (d), (e) and (f).}
\end{figure}

%\pspicture(0,0.2)(5.5,12.4)
%\rput[tl]{0}(-0.2,12.2){\epsfxsize=7.5cm \epsfclipon
%\epsffile{f1.eps}}
%\rput[tl]{0}(0,0.){
%%\rput[tl]{0}(0,2.25){
%\begin{minipage}{8.5cm}
%\vspace{1.5in}
%\small\parindent=3.5mm
%{\sc Fig.}~1.---
%Each plot shows the input primordial power spectrum (dashed curve), and mean power spectrum reconstructed from simulated $C_l$ data together with 1$\sigma$ uncertainties (solid curves) and the 5$\sigma$ confidence contours (dot-dashed curves). These are shown for the (a) scale-invariant, (b) saw-tooth and (c) Gaussian dip cases.
%\end{minipage}
%}
%\endpspicture


\begin{thebibliography}{}

\bibitem[Adams, Ross, \& Sarkar(1997)]{Adams97}
Adams, J.A., Ross, G.G., \& Sarkar, S. 1997, Nuclear Physics B, 503, 405

\bibitem[Adams, Cresswell, \& Easther(2001)]{Adams01}
Adams, J., Cresswell, B, \& Easther, R. 2001, Phys. Rev. D64, 3514

\bibitem[Barger, Lee, \& Marfatia(2003)]{Barger03}
Barger, V., Lee, H.S., \& Marfatia, D. 2003, astro-ph/0302150

\bibitem[Bennett et al.(2003)]{Bennett03}
Bennett, C., et al. 2003, astro-ph/0302207 

\bibitem[Bridle et al.(2003)]{Bridle03}
Bridle, S.L., Lewis, A.M., Weller, J., \& Efstathiou, G. 2003,
astro-ph/0302306

\bibitem[Burgess et al. (2005)]{Burgess05}
Burgess, C.P., Easther, R., Mazumdar, A., Mota, D.F., \& Multamaki, T., 2005, hep-th/0501125

\bibitem[Chung et al.(2000)]{Chung00}
Chung, D.J.H., Kolb, E.W., Riotto, A., \& Tkachev, I.I. 2000,
Phys. Rev. D, 62, 043508

\bibitem[Daubechies(1992)]{Daub92}
Daubechies, I. 1992, Ten Lectures on Wavelets, S.I.A.M., Philadelphia.

\bibitem[Elgaroy, Gramann, \& Lahav(2002)]{Elgaroy02}
Elgaroy, O.; Gramann, M.; Lahav, O. 2002
MNRAS, 333, 93
 
\bibitem[Feng \& Zhang(2003)]{FengZhang03}
Feng, B. \& Zhang, X. 2003, astro-ph/0305020

\bibitem[Hannestad(2004)]{hannestad04}
Hannestad, S 2004, JCAP, 0404, 002

\bibitem[Holman et al.(1991ab)]{Holman91ab}
Holman, R., Kolb, E.W., Vadas, S.L., \& Wang, Y. 1991a, Phys.\ Rev., D43, 3833

\bibitem[Holman et al.(1991b)]{Holman91b}
Holman, R., Kolb, E.W., Vadas, S.L., \& Wang, Y. 1991b, Phys.\ Lett., B269, 252

\bibitem[Hu \& Okamoto(2004)]{Hu04}
Hu, W., \& Okamoto, T. 2004, Phys. Rev. D69, 043004

\bibitem[Kogo, Sasaki \& Yokoyama(2004)]{kogo04}
Kogo, N, Sasaki, M, \& Yokoyama, J 2004, 

\bibitem[Lesgourgues(2000)]{lesgourgues00}
Lesgourgues, J 2000, Nucl. Phys. B, 582, 593

\bibitem[Lyth, Ungarelli, \& Wands(2002)]{Lyth02}
Lyth, D. H., Ungarelli, C., \& Wands, D. 2002, Phys. Rev. D, in press, astro-ph
/0208055

\bibitem[Matsumiya, Sasaki \& Yokoyama(2002)]{matsumiya02}
Matsumiya, M, Sasaki, M, \& Yokoyama, J 2002, Phys.Rev. D, 65, 083007; 
Matsumiya, M, Sasaki, M, \& Yokoyama, J 2002, JCAP 0302, 003
 
\bibitem[Mukherjee \& Wang(2003a)]{pia03}
Mukherjee, P., \& Wang, Y. 2003a, ApJ, 593, 38

\bibitem[Mukherjee \& Wang(2003b)]{pia03b}
Mukherjee, P., Wang, Y., et al. 2003b, ApJ, 599, 1

\bibitem[Mukherjee \& Wang(2003c)]{MW03c}
Mukherjee, P., Wang, Y., et al. 2003c, ApJ, 598, 779

\bibitem[Peiris et al.(2003)]{Peiris03}
Peiris, H. V., et al. 2003, astro-ph/0302225

\bibitem[Press et al.(1994)]{Press94}
Press, W.H., Teukolsky, S.A., Vettering, W.T., \& Flannery, 
B.P. 1994, Numerical Recipes, Cambridge University Press, Cambridge.

\bibitem[Randall, Soljacic, \& Guth(1996)]{Randall96}
Randall, L., Soljacic, M., \& Guth, A. 1996, 
Nucl. Phys. B472,  377

\bibitem[Seljak, McDonald \& Makarov(2003)]{Seljak03}
Seljak, U., McDonald, P., \& Makarov, A. 2003, astro-ph/0302571 

\bibitem[Shafieloo \& Souradeep(2004)]{SS04}
Shafieloo, A, \& Souradeep, T 2004, Phys.Rev. D, 70, 043523

\bibitem[Spergel et al.(2003)]{Spergel03}
Spergel, D. N.; et al. 2003, astro-ph/0302209 

\bibitem[Tegmark et al.2000]{TEHC}
Tegmark, M., Eisenstein, D.J., Hu, W. \& de Oliveira-Costa, A. 2000, ApJ, 530, 133

\bibitem[Tegmark, \& Zaldarriaga(2002)]{Tegmark02}
Tegmark, M., \& Zaldarriaga, M. 2002,
Phys.Rev. D, 66, 103508

\bibitem[Tocchini-Valentini, Douspis \& Silk(2004)]{valentini04}
Tocchini-Valentini, D, Douspis, M, \& Silk, J 2004, astro-ph/0402583

\bibitem[Wang(1994)]{Wang94}
Wang, Y. 1994, Phys.\ Rev., D50, 6135

\bibitem[Wang, Spergel, \& Strauss(1999)]{Wang99}
Wang, Y., Spergel, D.N., \& Strauss, M.A. 1999,
     ApJ, 510, 20
                         
\bibitem[Wang \& Mathews(2002)]{WangMathews02}
Wang, Y., \& Mathews, G.J. 2001, ApJ, 573,  1

\end{thebibliography}
\end{document}